\documentclass[preprint,showpacs,preprintnumbers,amsmath,amssymb, showkeys,superscriptaddress,titlepage]{revtex4-2}
\usepackage{graphicx}
\usepackage{dcolumn}
\usepackage{bm}
\usepackage{hyperref}
\usepackage{textcomp}

\begin{document}
\title{Production of charged kaons in $Ar Sc$ collisions.}
 
\author{G.I.~Lykasov}
\email{lykasov@jinr.ru}
\affiliation{Joint Institute for Nuclear Research, Joliot-Curie 6, 141980 Dubna, Russian Federation}

\author{A.I.~Malakhov}
\affiliation{Joint Institute for Nuclear Research, Joliot-Curie 6, 141980 Dubna, Russian Federation}
\affiliation{Dubna State University, Dubna, Moscow region, Russian Federation}

\author{A.A.~Zaitsev}
\affiliation{Joint Institute for Nuclear Research, Joliot-Curie 6, 141980 Dubna, Russian Federation}
\affiliation{Lebedev Physical Institute of the Russian Academy of Sciences (LPI RAS), 117997, Moscow, Russian Federation}

\begin{abstract}
  Inclusive spectra of the kaons produced in $Ar Sc$ collisions as
  functions of their transverse momentum $p_T$ at mid-rapidity have been
  calculated
  within the approach based on the assumption of the similarity of
  inclusive spectra of the hadrons produced in nucleus-nucleus collisions at
  their small transverse momenta in the mid-rapidity region taking into account
  the quark-gluon dynamics in nucleon-nucleon interactions. 
  This article gives a satisfactory description of NA61/SHINE data on $p_T$-spectra of the $K^\pm$
  mesons produced in $ArSc$ collisions at the mid-rapdity ($y\simeq$ 0).
\end{abstract}

\maketitle

\section{Introduction}
\label{intro}
A description of any physical observables, obtained in the collider
    experiments at LHC, is mainly based on different factorization theorems
    in quantum chromodynamics (QCD).
    According to these theorems, hard partonic processes are separated 
    from partonic density functions (PDF). An appropriate QCD evolution
    describing
    the PDF as a function of the resolution scale squared $\mu^2$,
    is given by the
    Dokshitzer-Gribov-Lipatov-Altarelli-Parisi (DGLAP) equation \cite{DGLAP}.  
    This approach can be successfully applied to analyze inclusive
    hard processes like deep--inelastic lepton-hadron scattering (DIS).      
    The transverse momentum ($k_T$) dependent (TMD) or unintegrated
    parton densities $f_a(x, {\mathbf k}_T^2, \mu^2)$ (with $a = q$ or $g$,
    x is the longitudinal momentum fraction of parton $a$ respective to the initial energy)
    satisfy the Balitsky-Fadin-Kuraev-Lipatov
    (BFKL) or Catani-Ciafaloni-Fioriani-Marchesini
    (CCFM) evolution equation \cite{CCFM}. However, these approaches obtained
    within the perturbative QCD have some difficulties to analyze the soft
    hadron production in $pp$ at low transfer $Q^2$, low transverse
    hadron
    momentum $p_T$ in the mid-rapidity ($y\simeq $ 0). It is due to the
    divergence of the QCD coupling constant $\alpha_s(Q^2)$.

    The relation of hard $pp$ processes and $ep$ DIS with soft hadron
    production in $pp$ collisions at low $p_t$ and mid-rapidity was offered
    in our paper \cite{GLLZ:2013}. It was shown that using the gluon
    distribution $f_g(x,k_T,Q^2)$  dependent of $x,Q^2$ and its transverse
    momentum
    $k_T$ (TMD) at
    low $Q^2$, which satisfies the
    saturation effect in the $ep$ DIS \cite{Mueller:1990,NZ:1990,GBW1,GBW2},
    one can describe
    satisfactorily the LHC data on inclusive $p_T$-spectra of charged hadrons
    produced in $pp$ collisions at low $p_T$ and the mid-rapidity region
    (soft hadron production). The saturation of the gluon TMD $f_g(x,k_T,Q^2)$
    means that at $Q$ less than the saturation scale $Q_s$ it does
    not depend on $Q^2$, see details also in
    \cite{LLM:2023,LLM2:2023,LLM3:2023}.  
    Therefore, the colliding protons at very low $Q^2$ can be presented as two
    systems consisting of three valence quarks and the gluon 
    environment. In our papers
    \cite{GLLZ:2013,LLM:2023,BGLP:2012} the
    hadron production in $pp$ collisions at LHC energies, at low $p_T$ and
    $y\simeq 0$ were calculated as the sum of the quark  and the
    gluon contributions. The similar procedure was used in \cite{ALM:2015,LM:2018,ML:2020} analyzing the
    pion and kaon production at middle and high energies. The NA61/SHINE data on
    inclusive $p_T$-data on $\pi^\pm$-, $K^\pm$-mesons production in $p p$ and
    $Be Be$
    collisions and ratios $\sigma_{K^\pm}/\sigma_{\pi^\pm}$ in the mid-rapidity as
    functions of $\sqrt{s}$ were successfully described using the gluon TMD
    mentioned above and a similarity principle of hadron inclusive spectra
    \cite{ALM:2015,LM:2018,ML:2020,LMZ:2021,LMZ:2022}. It was shown that the
    cross
    section ratios of kaons to pions produced in collisions of nuclei lighter
    than
    $Pb$ and $Au$, as a function of the initial energy
    $\sqrt{s_{NN}}$, in particular, $Be Be$ 
    and $Ar Sc$ have no peak,
    as revealed by the NA61/SHINE Collaboration
    \cite{NA61_BeBe:2021,NA61_ArSc:2021,ArSc:2023}.
    The fast increase of this 
    ratio, when $\sqrt{s_{NN}}$ grows from the kaon threshold up to 20-30 GeV
    and
    the slow increase at higher energies have been observed 
    \cite{NA61_BeBe:2021,NA61_ArSc:2021}.  
    Moreover, the energy dependence of $K/\pi$ ratios observed 
    in $BeBe$ collisions \cite{NA61_BeBe:2021} is similar to the one 
    in $pp$ collisions.

    The observation of a sharp peak in the production ratio
    of $K^+$ mesons  to $\pi^+$ mesons in central $PbPb$ collisions
    at mid-rapidity was done by the NA49 Collaboration
    \cite{NA49:2002,NA49:2008}.
    When the initial energy $\sqrt{s_{NN}}$ per nucleon becomes higher than
    30 GeV
    this ratio falls down.
    According to the assumption of \cite{Marek:1999,Marek:2014} this peak
    (so-called ``horn'') can appear due to formation of the quark-gluon plasma
    (QGP) phase at the center-of-mass energy $\sqrt{s}\simeq$ 7 GeV.
    However, there are other explanations of this ``horn'', see, for example,
    references in \cite{LMZ:2022}.
    
    In this paper we analyze the production of kaons in central $Ar Sc$ collisions
    at initial energies starting from the kaon threshold up to the LHC energy in the
    mid-rapidity region. In Section 2 we present the procedure to
    calculate inclusive spectra in $N N$ and $A A$, Section 3 consists of the calculation
    of inclusive $p_T$- spectra of $K^\pm$
    and their comparison to the NA61/SHINE data.

\section{The inclusive $p_T$-spectra of kaons for $NN$ and $AA$
      collisions at low transverse momenta and mid-rapidity.}

    As it is mentioned above, the perturbative QCD has some difficulties to
    analyze the
    soft hadron production in $pp$ collisions at low $Q^2$, small $p_T$ in the
    mid-rapidity region. It is due to the divergence of the QCD constant
    $\alpha_s(Q^2)$. On the other hand, the quark-gluon string model (QGSM)
    \cite{Kaid1,LAS} can be successfully applied to describe observables of hadron
    production in $pp$ collisions at large hadron momentum fractions $x$ and
    it is unable to analyze inclusive hadron spectra at $x\simeq 0$ or at
    the hadron rapidity $y\simeq $ 0.
    The application
    of the QGSM to calculate inclusive $p_T$-spectra of hadrons
    in $pp$ collisions at the $y=$ 0 and not large $p_T$ was done in
    \cite{BGLP:2012}
    assuming the contribution of nonperturbative gluon fragmentation into charged
    hadrons. This gluon contribution was parameterized as a function of
    $p_T$ fitting
    the LHC data on soft hadron spectra. Later in
    \cite{GLLZ:2013,LLM:2023,LLM2:2023,LLM3:2023} the gluon contribution
    to soft hadron production in $pp$ collisions at LHC energies was related
    to the
    saturation effect in the $ep$ DIS \cite{GBW1,GBW2}

    The inclusive hadron spectrum in nucleon-nucleon ($NN$) collisions
    at $y=$0 was
     $\rho_{NN}(y=0,p_T)$ presented in the  ,
    following form \cite{GLLZ:2013}: 
\begin{equation}
\rho_{NN}(y=0,p_T)~=~\rho_{q}(y=0,p_T)~+~\rho_{g}(y=0,p_T)~,
\label{def:rhoNN}  
\end{equation}
\begin{equation}
\rho_q(0,p_T)~=~g(s/s_00)^\Delta\phi_q(0,p_T)~;~
\label{def:rhoq}
\end{equation}
\begin{equation}
\rho_g(0,p_T)~=~(g(s/s_00)^\Delta-\sigma_{nd})\phi_g(0,p_T),
\label{def:rhog}                                                                
\end{equation}
where
\begin{equation}
\phi_q(0,p_T)~=~A_qexp(-b_q p_T)~,
\label{def:phi_q}
\end{equation}
\begin{equation}
\phi_g(0,p_T)~=~\sqrt{p_T}A_gexp(-b_g p_T),
\label{def:phi_g}                                                                
\end{equation}
Here $\rho_{q}(y=0,p_T)$ is the quark contribution and $\rho_{g}(y=0,p_T)$ is
the gluon one to the inclusive spectrum $\rho_{NN}(y=0,p_T)$ of hadrons produced
in nucleon-nucleon collisions.
The parameters were fitted in \cite{BGLP:2012} from the LHC data on soft hadron
production at $y=0$. 

We turn now to investigation of hadron production in $pp$ and $AA$ collisions at
mid-rapidity. In \cite{ALM:2015,LM:2018,ML:2020} a similar form for inclusive
hadron spectrum in $pp$ collisions as in Eq.~(\ref{def:rhoNN}) is presented.
Usually, the inclusive
$p_T$-spectrum $\rho_{pp}(y=0,p_T)$ is presented in the factorized form, a
function
of the initial energy $\sqrt{s}$ and the transverse momentum $p_T$, that is
valid at
high energies. However, at not high energies, especially at the energy close
to the
hadron threshold this factorization is broken, as it was shown firstly in
\cite{c3,4,c4} if one uses the four-momentum velocities of initial and final
particles. Analyzing the inclusive production of the cumulative hadrons produced in
$A A$ collisions, i.e., the production of hadrons  forbidden for free $NN$
collisions, and assuming the similarity of their inclusive spectra the parameter
similarity $\Pi$ was introduced in \cite{4}. It also was assumed that the
form of inclusive spectrum of the hadrons produced in $AA$ collisions depends
on $\Pi$. In fact, $\Pi$ is a function of $s,p_T,y$, it was calculated
analytically at $y=0$ in
\cite{5}, see also \cite{c4} and references therein, using the conservation law
of four-momenta in the inclusive reaction
:$A ~+~B \rightarrow~h~+~X$:
\begin{equation}
{(N_AP_A + N_BP_{B} - p_1)}^2 = 
{(N_Am_0 + N_B m_0 + M)}^2 ,
\label{eq:n2}
\end{equation}
where $N_A$ and $N_B$ are the fractions of the four-momentum transmitted by
nucleus $A$ and nucleus $B$, the forms of $N_A, N_B$  are presented in
\cite{LM:2018,5}; 
$P_A$, $P_B$, $p_1$ are the four-momenta of nuclei  
$A$,  $B$ and hadron $h$, respectively; $m_0$ is the mass of the nucleon; $M$ is
the mass of the 
particle providing conservation of the baryon 
number, strangeness and other quantum numbers.  
It allows us to find the minimal value of $M$,
which provides the conservation of quantum numbers.   
For $\pi$-mesons $m_1 = m_\pi$  and $M = $0.
For antinuclei $M=m_1$ and for $K^-$-mesons  $M = m_1 = m_K$,
$m_K$ is the mass of the $K$-meson.
For nuclear fragments $M = - m_1$.
For $K^+$-mesons $m_1 = m_K$ and $M = m_\Lambda  - m_0$,
$m_\Lambda$ is the mass of the $\Lambda$-baryon.
Let us note that the isospin effects of the produced hadrons and other nuclear
effects
are out of this approach. Therefore, it is assumed that within the similarity
approach there is no big
difference between the inclusive spectra of the $\pi^+$ and $\pi^-$ mesons
produced in $pp$ and $AA$
collisions. However, there is a difference between similar spectra of $K^+$
and $K^-$ mesons, because 
the values of $M$ are different. This is due to the conservation law of
strangeness.
In \cite{LM:2018,ML:2020,LMZ:2021,LMZ:2022} this approach was applied to
calculate inclusive $p_T$-spectra of pions and kaons produced in $p p$ and $Be Be$
collisions
and their cross section ratios, as functions of $\sqrt{s}$ of pions and kaons
produced. It allowed us to describe the
data of the NA61/SHINE Collaboration and the data obtained by
RHIC, BNL and LHC
rather
successfully. 
 In \cite{4,5} the parameter of self-similarity is introduced in the following 
form:
\begin{equation} 
\Pi=\min \frac{1}{2} \left[ (u_A N_A + u_B N_B)^2\right]^{1/2}  ,
\label{eq:n3} 
\end{equation}                                            
where $u_A$ and $u_B$ are the four-velocities of nuclei $A$ and $B$, respectively. 
The minimization over 
$N$ presented in Eq.~(\ref{eq:n3}) allows us to find the parameter $\Pi$. This
parameter introduced in \cite{4} was obtained as the analytical form in \cite{5}
for nucleus-nucleus
collisions in the mid-rapidity region. Thus, it can
also be applied successfully for the analysis of 
pion production in $pp$ collisions, as it was shown in
\cite{ALM:2015,LM:2018,ML:2020}. It was offered in \cite{LM:2018} to use the
four-momentum velocities the inclusive hadron spectrum in $pp$ collisions
$\rho_{pp}(y=0,p_T)$ given in Eq.~\ref{def:rhoNN} can be presented as a function
of the similarity parameter of $\Pi$, which at high energies $\sqrt{s}$ goes to
the transverse momentum $p_T$ at high energies $\sqrt{s}$ and $y=0$.

The inclusive spectrum of hadron $h$ produced in the $AB$ collision
can be parameterized as a general universal function dependent on the 
similarity parameter $\Pi$, as it was shown in \cite{Baldin_AA:1996}:
\begin{equation}
E d^3 \sigma_{AB}/d^3p~=~A_A^{\alpha(N_A)}\cdot A_B^{\alpha(N_{B})}\cdot F(\Pi)
\label{eq:n4} 
\end{equation}
where $\alpha(N_A)=1/3 + N_A/3$, $\alpha(N_B)=1/3 + N_B/3$ 
and function $F(\Pi)$ is the inclusive spectrum of hadron production in the $A B$
collision. Here $F(\Pi)$ at $x=$ 0 has the same form as $\rho_{NN}(x=0,p_t)$ presented
in Eq.~\ref{def:rhoNN} with substitution of transverse momentum $p_T$ by $\Pi$
\cite{LM:2018,LMZ:2021}. 
     
At $y=$ 0 $N_A=N_B=N$ the function $\Pi$ is found from 
the minimization of Eq.~\ref{eq:n3} by solving the equation
\cite{4,5}:
\begin{eqnarray}
\frac{d\Pi}{d N}=0
\label{def:minimPi} 
\end{eqnarray}  
The exact solution of Eq.~\ref{def:minimPi} at $y=0$,
as 
\begin{eqnarray}
N=\frac{\Pi}{\mbox{cosh}(Y)}\equiv\frac{2m_0\Pi}{\sqrt{s}}, 
\label{def:N} 
\end{eqnarray}  
was obtained in \cite{5}, for details see, also \cite{LM:2018}. In Eq.~\ref{def:N} $Y$ is rapidity of colliding nuclei.  
Therefore, $\alpha(N)=1/3 + 2m_0\Pi/(3\sqrt{s})$.
Function $F(\Pi)$ has the following form \cite{LM:2018}:
\begin{eqnarray}
F(\Pi)=\bigg[ A_q \mbox{exp}\Big(-\frac{\Pi}{C_q}\Big) +
A_g\sqrt{p_T}\phi_1(s) \mbox{exp}\Big(-\frac{\Pi}{C_g}\Big)\bigg] \sigma_{tot}
\label{def:F} 
\end{eqnarray}
where 
\begin{eqnarray}
\Pi(s,m_{1T},y)~=~\left\{\frac{m_{1T}}{2m_0\delta_h}+
\frac{M}{\sqrt{s}\delta_h}\right\}\mbox{cosh}(y)G,
\label{eq:n10}
\end{eqnarray}
\begin{eqnarray}
G~=~\left\{1+\sqrt{1+\frac{M^2-m_1^2}
{(m_{1T}+2Mm_0/\sqrt{s})^2\mbox{cosh}^2(y)}\delta_h}\right\}~
\label{def:G}.
\end{eqnarray}
Here 
$\phi_1(s)~=~1-\sigma_{nd}(s)/\sigma_{tot}(s)$, see \cite{LM:2018,ML:2020},
$\delta_h=\left(1 - \frac{s_{th}^h}{s} \right)$;
$s_{th}^{\pi}\simeq 4m_0^2$;
$s_{th}^{K^+}=\left(m_0 + m_K + m_\Lambda \right)^2$;
$s_{th}^{K^-}=(2m_0+2m_K)^2$;
$M = m_\Lambda - m_0; m_\Lambda = $ 1.115 GeV; 
$m_K = $ 0.494 GeV; $s_0 = $ 1 GeV; $m_0 = 0.938$ GeV;
$p_{1T}$ and $m_{1T}$ are the transverse momentum and transverse mass
of the produced hadron $1$; 
$\sigma_{nd} = (\sigma_{tot} - \sigma_{el} - \sigma_{SD})$ is the 
non-diffractive cross-section;
$\sigma_{tot},\sigma_{SD}$ and $\sigma_{el}$ are the total
cross-section, the single diffractive cross-section and the elastic
cross-section of $pp$ collisions, respectively. They were taken from
\cite{sigma:2013} and \cite{sigm_el:2017} and, together with
 parameters $A_q, C_q$ and  $A_g, C_g$, they are presented in the Appendix.
 

\section{Results and discussion}

In the case of the process $AB\rightarrow h + X$ Eq.~\ref{eq:n4} looks
as follows:
\begin{eqnarray}
\rho_{AB}^h(p_{hT},y)~\equiv E_h\frac{d^3 \sigma^h_{AB}}{d^3p_1}~=
\frac{1}{\pi}\frac{d\sigma^h_{AB}}{dp_{1T}^2dy}~= \\
\nonumber
\frac{1}{\pi}\frac{d\sigma^h_{AB}}{dm_{1T}^2dy}=(AB)^{\alpha(N)}
F(\Pi(s,m_{1T},y)) ,
\label{eq:AA_yspectrum}
\end{eqnarray}
Then, the production cross-section of  hadron $h$ in $AA$ collisions
integrated over its
transverse momentum $p_{1T}$ or transverse mass $m_{1T}$ at $y=0$ and $s\geq s_{th}^h$ can be presented in the following form:
\begin{eqnarray}
	\frac{d\sigma^h_{AB}}{dy}(s,y=0)	=2\pi\int_{p^{\rm{min}}_{1T}}^{p^{\rm{max}}_{1T}}\rho_{h_1}^{AB}(s,p_{1T},y=0)p_{1T}dp_{1T}
\end{eqnarray}
  
\begin{figure}
	\begin{center}
		\includegraphics[width=10cm]{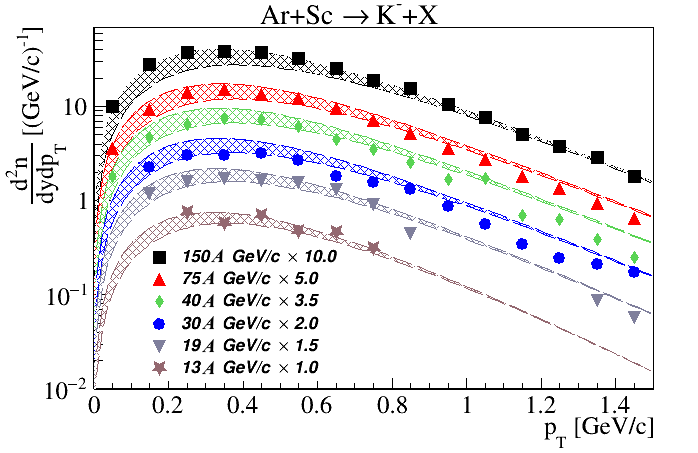}

	\caption{Inclusive $p_T$-spectra of $K^-$-mesons produced in 10\% most central $ArSc$
	  collisions at mid-rapidity ($y\approx$0) and different initial
          momenta. NA61/SHINE data were taken from \cite{ArSc:2023}.
	}
	\label{figKmin}
\end{center}
\end{figure}
 
\begin{figure}
\begin{center}
  \includegraphics[width=10cm]{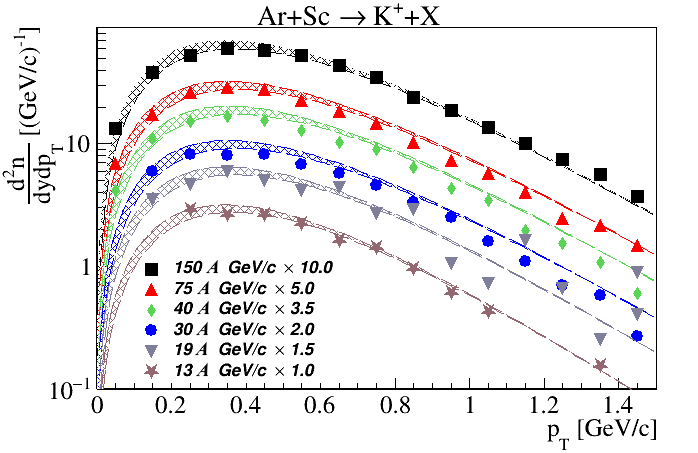}
  \caption{Inclusive $p_T$-spectra of $K^+$-mesons produced in 10\% most central $ArSc$
    collisions at mid-rapidity ($y\approx$0) and different initial momenta. NA61/SHINE data were taken from \cite{ArSc:2023}.
  }
\label{figKpl}
\end{center}
\end{figure}

The $p_T$ spectra of $K^+$ and $K^-$ mesons are the sums
of quark and gluon contributions including uncertainties due to the fit of data 
shown in Figs.~(\ref{figKmin},\ref{figKpl}). 
By fitting NA61/SHINE data \cite{ArSc:2023} on $p_T$ spectra at mid-rapidity the parameters
$C_q, A_g$ and $C_g$ were found to be independent of the initial energy $\sqrt{s}$,
they depend on the kind of produced mesons $K^+,K^-$.
However, the parameter $A_q$ varies very weakly at initial momenta from
13$A$ GeV$/$c up to 150$A$ GeV$/$c. In Figs.~(\ref{figKmin},\ref{figKpl}) our calculations are presented with two curves (the upper and bottom) for each energy.  All the bottom lines for the negatively charged kaons have been calculated with the parameter $A_q=$ 2.56$\pm$0.22 
corresponding to satisfactory description of the data at $P_{in}=$ 30$A$ GeV$/$c (Fig.~\ref{figKmin}). 
The upper lines in Fig.~\ref{figKmin} belong to the upper curve at $P_{in}=$ 150$A$ GeV/c calculated with parameter $A_q=$ 5.43$\pm$0.08.
At the same time, all upper lines in Fig.~\ref{figKpl} are the result of calculations with the parameter $A_q=$ 1.6$\pm$0.3 given while fitting the data at 13$A$ GeV$/$c. The lower lines correspond to the calculations with minimal parameter of $A_q=$ 0.72$\pm$0.22 extracted from fitting of data at 30$A$ GeV$/$c. The set of obtained parameters  $A_q, C_q, A_g$ and $C_g$ is given in Table 1.
The uncertainties in $p_T$ spectra are due to the uncertainties of the parameter $A_q$.

\section{Conclusion}

In this paper we have applied the similarity approach, based on the assumption of the similarity of inclusive spectra of the
hadrons produced 
in $AA$ collisions at their low transverse momenta and in the mid-rapidity region. 
The essential feature of our modification of this approach is as follows. We propose to include the quark-gluon dynamics of hadron
production in nucleon-nucleon interaction at mid-rapidity. It leads to a more complicated 
form of the $p_T$ spectrum of hadrons produced in $NN$ interaction compared to the simple exponential form used 
in the similarity approach \cite{4,c4}. Moreover, the correct Regge asymptotic behavior of the total and inelastic $NN$ 
cross-sections is taken into account. It results in the realistic energy dependence of inclusive $p_T$ spectra of hadrons 
produced in $pp$ and $AA$ collisions at not large values of $p_T$ in the mid-rapidity region. Within this approach
we have analyzed the 
production of charged kaons in $ArSc$ collisions at mid-rapidity 
0 $< y < 0.2$ within a wide range of initial energies.
We have presented a self-consistent satisfactory description
of the NA61/SHINE data on $p_T$-spectra of
kaons at 5.12 GeV $\leq\sqrt{s}\leq$ 16.84 GeV.
Note, that the approach used in this paper is valid at the mid-rapidity region
mentioned above. However, the NA61/SHINE data
are given in \cite{ArSc:2023} at
$y>0.6$, therefore we have not calculated the inclusive spectra of pions
and the ratio of cross sections of $K$-mesons to $\pi$-mesons.

\section{Appendix} 
The parameterizations of $\sigma_{tot},\sigma_{SD}$ and $\sigma_{el}$ 
have the following forms \cite{sigma:2013} and \cite{sigm_el:2017}: \\
$\sigma_{tot} = (21.7(s/s_0)^{0.0808} + 56.08(s/s_0)^{-0.4525}$) mb;\\ 
$\sigma_{el} = (12.7 - 1.75\mbox{ln}(s/s_0) + 0.14\mbox{ln}^2(s/s_0)$) mb;\\ $\sigma_{SD} = (4.2 +
\mbox{ln}(\sqrt{s/s_0})$) mb.

In Figs.~(\ref{figKmin},\ref{figKpl}) the $p_T$-spectra of the charged
kaons, produced in the mid-rapidity of central $ArSc$ collisions within the
initial momentum range of (13-150)$A$ GeV/c \cite{ArSc:2023}, obtained by
the NA61/SHINE Collaboration, 
are illustrated together with our calculations.
The parameters $A_q,A_g$ and $C_q,C_g$ were found from
a fit of NA61/SHINE data and are shown 
in Table 1. Parameters $A_q$ for $K^+$ and $K^-$ meson
production were found from the fit of NA61/SHINE data
at initial energies $P_{in}=$ (30-150)$A$ GeV$/$c. 
Parameters $A_g$ for $K^+$ and $K^-$ meson
production were obtained from the fit of NA61/SHINE data
at $P_{in}=$ 150$A$ GeV$/$c. 
Other parameters $C_q$ and $C_g$ were taken from fits of NA61/SHINE data in
$pp$ collisions.   
\begin{table*}[h]
\centering
\caption{Parameters found from the fit of NA61/SHINE data.}
\begin{tabular}{ | c | c | c | c | c | c | c | c| c | c | c | c | c |}
\hline
Ar+Sc$ \to h+X$  & \multicolumn{6}{c|}{$K^{+}$} & \multicolumn{6}{c|}{$K^{-}$}   \\ \hline
$\sqrt{s}$, $A$ GeV & 16.8 & 11.9 & 8.8 & 7.6 & 6.1 & 5.1 & 16.8 & 11.9 & 8.8 & 7.6 & 6.1 & 5.1 \\ \hline
$\textit{P}_{in}$, $A$ GeV/$c$ & 150 & 75 & 40 & 30 & 19 & 13 & 150 & 75 & 40 & 30 & 19 & 13 \\ \hline
A$_{q}$ & 1.05 & 1.17 & 0.82 & 0.72 & 1.172 & 1.574 & 5.43 & 4.098 & 3.22 & 2.56 & 3.53 & 4.74 \\ \hline
C$_{q}$ & \multicolumn{12}{c|}{0.148}\\ \hline
A$_{g}$ & \multicolumn{6}{c|}{7.72} & \multicolumn{6}{c|}{2.561}\\ \hline
C$_{g}$ & \multicolumn{6}{c|}{0.2066} & \multicolumn{6}{c|}{0.2271}\\ 
\hline
\end{tabular}
\label{table 1}
\end{table*}


{\bf Acknowledgements.}

\begin{sloppypar} 
  We are very grateful to M. Gazdzicki for extremely helpful discussions.   
\end{sloppypar}


\begin{thebibliography}{00}
\bibitem{DGLAP}
  V.N. Gribov{,} L.N.~Lipatov{,} Sov. J. Nucl. Phys. {\bf 15}{,} 438 (1972); \\
  L.N.~Lipatov{,} Sov. J. Nucl. Phys. {\bf 20}{,} 94 (1975); \\ G.~Altarelli{,}
  G.~Parisi{,} Nucl. Phys. B {\bf 126}{,} 298 (1977); \\ Yu.L. Dokshitzer{,}
  Sov. Phys. JETP {\bf 46}{,}~641 (1977).

\bibitem{CCFM}
M.~Ciafaloni{,} Nucl. Phys. B {\bf 296}{,} 49 (1988); \\ S. Catani{,} F.
  Fiorani{,} G. Marchesini{,} Phys. Lett. B {\bf 234}{,} 339 (1990); \\ S.
  Catani{,} F. Fiorani{,} G. Marchesini{,} Nucl. Phys. B {\bf 336}{,} 18
  (1990); \\ G. Marchesini{,} Nucl. Phys. B {\bf 445}{,}~49 (1995).

\bibitem{GLLZ:2013} {A.A.~Grinyuk, G.I.~Lykasov, A.V.~Lipatov, N.P.~Zotov}, 
Phys.Rev. \textbf{D87}, 074017 (2013).


\bibitem{Mueller:1990}
A.H.~Mueller, Nucl.Phys., B {\bf 335}, 115 (1990).  

\bibitem{NZ:1990}
N.~Nikolaev, B.G.~Zakharov, Z.Phys. C{\bf 49}, 607 (1990).

\bibitem{GBW1}
K.~Golec-Biernat, M.~Wuestoff, Phys.Rev.D {\bf 59}, 014017 (1998).

\bibitem{GBW2}
K.~Golec-Biernat, M.~Wuestoff, Phys.Rev.D {\bf 60}, 114023 (1999).  

\bibitem{LLM:2023}
A.V.~Lipatov, G.I.~Lykasov, M.A.~Malyshev,
Phys.Rev. D {\bf 107}, 014022 (2023).

\bibitem{LLM2:2023}
A.V.~Lipatov, G.I.~Lykasov, M.A.~Malyshev,
Phys.Lett.B {\bf 839}, 137780 (2023).

\bibitem{LLM3:2023}
A.V.~Lipatov, G.I.~Lykasov, M.A.~Malyshev,
Phys.Lett.B {\bf 848}, 137390 (2024).

\bibitem{BGLP:2012}
V.A.~Bednyakov, A.A.~Grinyuk, G.I.~Lykasov, M.~Pogosyan,
Int.J.Mod.Phys., \textbf{A27}, 1250042 (2012). 

\bibitem{ALM:2015}
D.A.~Artemenkov, G.I.~Lykasov, A.I.~Malakhov,
Int.J.Mod.Phys. \textbf{A30}, 1550127 (2015).

\bibitem{LM:2018}
G.I.~Lykasov, A.I.~Malakhov, Eur. Phys. J. A {\bf 54}, 187 (2018).

\bibitem{ML:2020}
A.I.~Malakhov, G.I.~Lykasov,  Eur. Phys. J. A {\bf 56}, 114 (2020).

\bibitem{LMZ:2021}
G.I.~Lykasov, A.I.~Malakhov, A.A.~Zaitsev, Eur. Phys. J. A {\bf 57 }, 78 (2021).

\bibitem{LMZ:2022}
G.I.~Lykasov, A.I.~Malakhov, A.A.~Zaitsev, Eur. Phys. J. A {\bf 58 }, 112
(2022).

\bibitem{NA61_BeBe:2021}
A.~Acharya, et al., (NA61/SHINE Collaboration) Eur. Phys. J. C {\bf 81}, 73
(2021).
 
\bibitem{NA61_ArSc:2021}
  A.~Acharya, et al., (NA61/SHINE Collaboration), Eur. Phys. J. C {\bf 81}, 397 (2021).

\bibitem{ArSc:2023}
  H. Adhikary, et al., (NA61/SHINE Collaboration) arXiv:2308.16683 [nucl-ex]

\bibitem{NA49:2002}
S.V.~Afanasiev, et al., (NA49 Collaboration) Phys.Rev.C \textbf{66}, 054902 (2002).

\bibitem{NA49:2008}
C.~Alt, et al.,  (NA49 Collaboration) Phys.Rev.C \textbf{77}, 024903 (2008).

\bibitem{Marek:1999}
M.~Gazdzicki, M.I.~Gorenstein, Acta Physika Polon., B {\bf 30}, 2705 (1999).

\bibitem{Marek:2014} 
  M.~Gazdzicki, M.I.~Gorenstein, P.~Seyboth, J.Mod.Phys. E {\bf 23}, 1430008 (2014).

\bibitem{Kaid1}
A.B.Kaidalov,
Z.Phys., {\bf C12}, 63 (1982); Sarveys High Energy Phys., {\bf 13}, 265 (1999).
A.B.Kaidalov, O.I.Piskunova, Z.Phys., C{\bf 30}, 145 (1986);
Yad.Fiz., {\bf 43}, 1545 (1986).

\bibitem{LAS}
G.I.Lykasov, G.G.Arakelyan, M.N.Sergeenko, EPAN,v.30,p.817 (1999).

\bibitem{c3}  
{A.M.~Baldin, L.A.~Didenko}, Fortsch.Phys. \textbf{38}, 261 (1990).

\bibitem{4}
A. M.Baldin, A. A. Baldin. Phys. Particles and Nuclei, {\bf 29} No3, 232 (1998).

\bibitem{c4}  
{A.M.~Baldin, A.I.~Malakhov, and A. N.~Sissakian}, Phys. Part. Nucl. \textbf{29}
(Suppl. 1), 4 (2001).

\bibitem{5}
A.M.~Baldin, A.I.~Malakhov. JINR Rapid Communications, No.1(87)-98, pp.5-12
(1998).

\bibitem{Baldin_AA:1996}
 A.A.~Baldin, JINR  Rapid  Comm. No. 4[78]-96  p.61-68.
\bibitem{sigma:2013}
N.Cartiglia, arXiv:1305.6131 [hep-ex]

\bibitem{sigm_el:2017}
 S.H.Stark,Eur.Phys.J. (Web of Conf.) {\bf 141}, 03007 (2017). 

\end{thebibliography}
\end{document}